\documentclass[amsmath,amssymb,aps,prd,10pt,twocolumn,showkeys]{revtex4}
\usepackage{graphicx}
\usepackage{mathtools}
\usepackage{verbatim}
\usepackage{placeins}
\usepackage{hyperref}

\begin{document}

\title{Complex Analysis of Askaryan Radiation: UHECR Reconstruction}

\author{Jordan C. Hanson}
\email{jhanson2@whittier.edu}
\affiliation{Department of Physics and Astronomy, Whittier College}
\author{Damian Ibá\~{n}ez-Rodriguéz}
\affiliation{Department of Physics and Astronomy, Whittier College}
\date{\today}

\begin{abstract}
Ultra-high energy cosmic rays (UHECR) can produce relativistic cascades that emit radio-frequency (RF) pulses in the 0.1-1 GHz bandwidth via two distinct effects: the geomagnetic effect, and the Askaryan effect.  The geomagnetic effect occurs when the magnetic field of the Earth causes cascade charges to form a transverse current that radiates linearly polarized radiation aligned with the Lorentz force direction.  The Askaryan effect is caused by the net negative charge excess in the cascade that radiates linearly polarized radiation along the Cherenkov cone.  When UHECR cascades enter solid, RF transparent matter at altitudes where the cascade develops, Askaryan radiation created in the solid medium can propagate to RF detectors.  The Askaryan Radio Array (ARA) at the South Pole has observed 13 UHECR candidates in precisely this fashion.  We present an analytical model that confirms the events are UHECRs.  The model includes Askaryan radiation created within the ice sheet and the ARA RF channel response.  The coherently summed waveforms (CSWs) from the UHECR candidates match our model with correlation coefficients between 0.69 and 0.86, and with minimal fractional power differences.  Finally, we demonstrate how to obtain the Askaryan $\vec{E}$-field from our model, and that it matches the results from the ARA collaboration.
\end{abstract}


\maketitle

\section{Introduction}
\label{sec:int}

UHECR cascades produce RF pulses via the geomagnetic and Askaryan effects, and these pulses have been used to study UHECR properties \cite{huege2017radiowave-6f6,schroder2025radio-14a}.  Detectors with RF channels sensitive to RF pulses from UHECRs add complementary data for event reconstruction to traditional ground-based muon and fluorescence detectors.  UHECR candidate events have now been presented by the Antarctic Ross Ice Shelf Antenna Neutrino Array (ARIANNA), ARA, and Radio Neutrino Observatory Greenland (RNO-G) collaborations \cite{barwick2017radio-c4a,xwqy-yzrk,agarwal2025validating-d6c}.  ARA is located at the South Pole, 2800 meters above sea level, where the thinner atmosphere allows for UHECR cascades to develop near its RF detection channels.  As charged particles from the UHECR cascade enter the ice, the negative excess charge radiates collectively in the 0.1-1 GHz bandwidth.  The Askaryan signal dominates the geomagnetic signal in vertically polarized RF channels at the South Pole, because the vertical orientation of the magnetic field of the Earth causes a horizontal polarization of the geomagnetic signal.

Within the genre of calculations that predict the properties of RF pulses from UHECRs and ultra-high energy neutrinos (UHE-$\nu$), there are three classes of models.  First, Monte Carlo simulations may be used to track each particle and corresponding RF emission within the cascade.  These are known as full Monte Carlo, or \textit{full-MC calculations} \cite{zas1992electromagnetic-a30,10.1016/j.astropartphys.2009.06.005,alameddine2025simulating-06f}.  Second, there are formalisms in which the analytic vector potential corresponding to the radiating charge is convolved with the simulated cascade profile.  These are called \textit{semi-analytic calculations}, and have been used to predict the RF pulses from UHE-$\nu$ interacting in Antarctic and Greenlandic ice \cite{10.1103/physrevd.84.103003,PhysRevD.101.083005,10.1140/epjc/s10052-020-7612-8}.  Finally, there are \textit{fully analytic} calculations that derive the UHECR or UHE-$\nu$ signal from classical electrodynamics.  Examples of such models are Ralston and Buniy (RB2001) \cite{10.1103/physrevd.65.016003}, Hanson and Connolly (HCRB2017) \cite{10.1016/j.astropartphys.2017.03.008}, and Hanson and Hartig (HH2022) \cite{PhysRevD.105.123019}.

One key advantage of fully analytic Askaryan calculations is that they can be fit to raw data from Askaryan-class detectors.  A prediction for the voltage trace from an RF detection channel is produced by convolving the analytic Askaryan model with an accurate model of the RF channel response.  Analytic voltage traces can be matched to the observed voltage traces by maximizing the correlation coefficient from cross-correlation while tuning parameters like the Askaryan pulse width.  Recently, Hanson and Hartig presented evidence that the HH2022 model, when convolved with an RF channel model, produces analytic voltage traces that match voltage traces from UHE-$\nu$ signals in NuRadioMC.  This technique can be used to reject several years of thermal noise backgrounds in detectors like ARA and RNO-G (HH2026) \cite{10.1140/epjc/s10052-020-7612-8,hanson2026complex-a31}.  Further, once the match between observed and analytic voltage trace is optimized, the Askaryan electromagnetic pulse can be extracted based on the fit parameters.  In this work, we show that the HH2026 procedure matches the voltage traces from 13 UHECR candidate events observed by ARA \cite{xwqy-yzrk}.

The remaining sections of this work are organized as follows: in Sec. \ref{sec:unit}, we define units, functions, operations on functions, and notational conventions.  In Sec. \ref{sec:ara}, we review the design of the Askaryan Radio Array detector, including the properties of the RF channels.  In Sec. \ref{sec:recon}, we present the match between our theoretical voltage traces and the UHECR candidate events.  We note that the UHECR candidate event data have been publicly released by the ARA Collaboration (\url{https://zenodo.org/records/19578836}).  Finally, in Sec. \ref{sec:conc}, we summarize our results and indicate future directions of the research.

\section{Units, Definitions, and Conventions}
\label{sec:unit}

Our goal is to demonstrate a match between a theoretical voltage trace for a UHECR candidate in RF detection channels and the observed voltage traces from those same channels.  Let the Askaryan \textit{signal} created by the portion of the UHECR cascade radiating in the ice be $s(t)$.  Let $E_0$ be the amplitude in V m$^{-1}$ ns$^{-1}$, $t$ be the time in ns relative to peak emission, and $\sigma_t$ be the pulse width in ns.  According to Eqs. 28 and 9 and from \cite{PhysRevD.105.123019,hanson2026complex-a31}, respectively, $s(t)$ is 

\begin{equation}
s(t) = -E_0 t e^{-\frac{1}{2}\left(t/\sigma_t\right)^2} \label{eq:s}
\end{equation}

To predict a voltage trace from RF dipole channels in ARA, $s(t)$ must be convolved with $r(t)$, the RF channel response.  Let $R_0$ be the amplitude of the response, with units of m ns$^{-1}$, $f_0$ be the resonance frequency of the dipole in GHz, and $\gamma$ be the exponential decay constant in GHz.  According to Eq. 31 from \cite{hanson2026complex-a31}, $r(t)$ is

\begin{equation}
r(t) = R_0 e^{-2 \pi \gamma t} \cos(2\pi f_0 t) \label{eq:r}
\end{equation}

The predicted voltage trace is given by the convolution of $s(t)$ and $r(t)$, denoted $s(t) * r(t)$.  The convolution is defined as:

\begin{equation}
s(t) * r(t) = \int_{-\infty}^{\infty} s(t-\tau) r(\tau) d\tau 
\end{equation}

Let $u(t)$ represent the Heaviside step function.  According to Eq. 55 from \cite{hanson2026complex-a31}, $s(t) * r(t)$ becomes

\begin{multline}
s * r = \\ -E_0 R_0 \int_{-\infty}^{\infty} (t-\tau) e^{-\frac{1}{2}\left(\frac{t-\tau}{\sigma_t}\right)^2} \Re\left\lbrace e^{2\pi j f_0 \tau} e^{-2\pi\gamma \tau} \right\rbrace u(\tau) d\tau \label{eq:sr}
\end{multline}

The Heaviside step function is introduced to ensure causality.  When the voltage trace is predicted with $s(t) * r(t)$, the result must be matched to observed traces from the RF channels with fixed, measured values for $\gamma$ and $f_0$.  The match is achieved by tuning $\sigma_t$ to maximize the peak value of the cross-correlation.  For real signals $f(t)$ and $g(t)$, cross-correlation is defined as

\begin{equation}
f(t) \star g(t) = \int_{-\infty}^{\infty} f(\tau) g(t+\tau) d\tau
\end{equation}

The algorithm used to compute the cross-correlation in this analysis is \verb+scipy.signal.correlate+.  To maximize the signal to noise ratio (SNR) of UHECR data, traces from multiple channels are combined to form the \textit{coherently summed waveform} (CSW). 

The CSW is computed from a set of traces using the following algorithm: (1) select a reference trace, (2) cross-correlate with the next trace in the set until the cross-correlation is maximized, (3) add the two traces to form a new reference, and (4) repeat until each trace in the set is added coherently to the reference.  The CSW is normalized so that the inner product of the trace is one.  The theoretical trace is normalized in the same way.  Normalized traces are unit-less but retain nanoseconds for the time unit.

The solution to Eq. \ref{eq:sr} was shown in Eq. 61 of \cite{hanson2026complex-a31}.  Let $j = \sqrt{-1}$, $x = t/(\sqrt{2}\sigma_t)$, $z = (2\pi j f_0 - 2\pi\gamma)\sqrt{2}\sigma_t$, $w(z)$ be the Faddeeva function of a complex variable, and

\begin{equation}
q = -j\left(x + \frac{z}{2}\right) \\
\end{equation}

The solution to Eq. \ref{eq:sr} is

\begin{multline}
s * r = -\sqrt{\pi}R_0 E_0 \sigma_t^2 \\ \left(xe^{-x^2} \Re\left\lbrace w(q) \right\rbrace - \frac{1}{2}e^{-x^2} \Re\left\lbrace -j \frac{d w(q)}{dq} \right\rbrace \right) \label{eq:sr2}
\end{multline}

A Python3 code was given in \cite{hanson2026complex-a31} to compute Eq. \ref{eq:sr2}.  Note that $q$ depends on $x$, which depends on $t$.  The $s * r$ oscillates at a frequency $\approx f_0$ because the Faddeeva function oscillates with a complex argument.  The envelope of $s * r$ is limited by the width of $s(t)$ and the exponential decay of $r(t)$.

\section{The Askaryan Radio Array}
\label{sec:ara}

\begin{figure}
\centering
\includegraphics[width=0.49\textwidth]{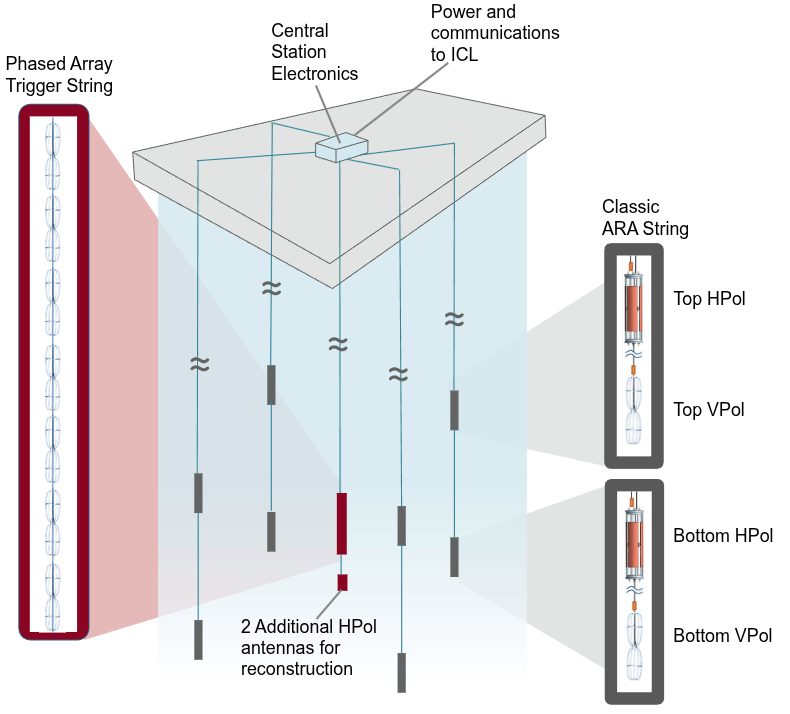}
\caption{\label{fig:ara} A general schematic of station A5 within the Askaryan Radio Array (ARA) detector.  The RF phased array trigger string is in the center, surrounded by four reconstruction strings.  Strings have both HPol and VPol RF channels.  Figure adopted from \cite{allison2022lowthreshold-795}.}
\end{figure}

The Askaryan Radio Array consists of five stations deployed near the South Pole designed to detect UHE-$\nu$ and UHECR events via the Askaryan effect in Antarctic ice \cite{allison2022lowthreshold-795,10.1103/physrevd.102.043021,allison2012design-cd3,testbed}.  Station A5 utilizes a dedicated RF trigger on the CSW output of a phased array of seven vertically polarized dipoles.  The RF phased array trigger boosts the SNR of UHE-$\nu$ and UHECR events, and allows for increased sensitivity to rare fluxes of these particles \cite{avva2017development-713}.  The ARA UHECR candidates used in this analysis were recorded on station A5 via the RF phased array trigger.  A schematic of the A5 detector is shown in Fig. \ref{fig:ara}, and the A5 detector is summarized in \cite{allison2022lowthreshold-795}.

\begin{figure}
\centering
\includegraphics[width=0.49\textwidth]{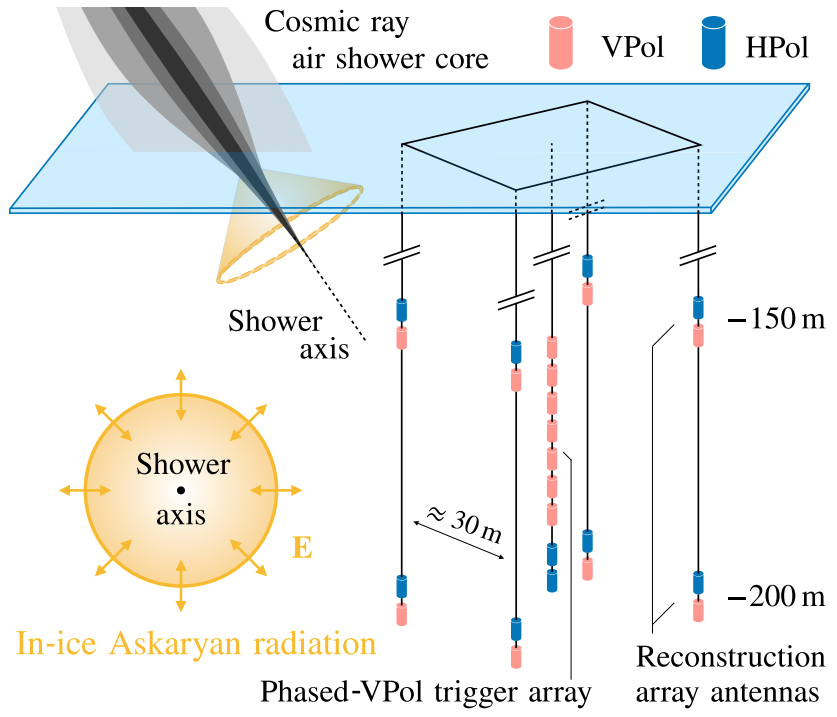}
\caption{\label{fig:ara2} The proposed UHECR candidate event geometry consists of a high-energy cascade interacting along the shower axis that enters the ice and creates Askaryan radiation.  Figure adopted from \cite{xwqy-yzrk}.}
\end{figure}

The UHECR candidate events were detected with the RF phased array trigger in A5.  The proposed event geometry is shown in Fig. \ref{fig:ara2}.  Typically UHECR candidates create both geomagnetic and Askaryan radiation, with geomagnetic radiation dominating if the detector lies far beneath the cascade and propagating at a latitude where the air shower axis forms a large angle with the local magnetic field. For example, the ARIANNA collaboration measured UHECR events, at a latitude of 78.7 degrees S and at an altitude of 576 m above sea level, that were geomagnetically dominated \cite{barwick2017radio-c4a,10.3189/2015jog14j214}.  Similar results have been shared by the RNO-G collaboration \cite{agarwal2025validating-d6c}.  

Although RNO-G and ARA are both situated at high altitudes, their respective latitudes of 72 degrees N and 90 degrees S play a key role in observable data.  For contrast, at 90 degrees South, geomagnetic radiation is expected to be horizontally polarized due to the vertical orientation of the local magnetic field.  The vertically polarized (VPol) RF channels are therefore insensitive to geomagnetic radiation.  Data from the horizontally polarized (HPol) RF channels has not been published and is not included in our analysis.  For the VPol channels, we will show in Sec. \ref{sec:recon} that geomagnetic radiation is not necessary to explain the observed voltage traces.

\section{Reconstruction Analysis}
\label{sec:recon}

Our cross-correlation analysis between Eq. \ref{eq:sr2} and the CSWs from the UHECR candidates from A5 begins by making two assumptions.  The first is that the $f_0$ and $\gamma$ parameters are identical across the A5 phased array trigger string.  Even though each channel may not have \textit{identical} parameters, the intent of the design shown in Fig. \ref{fig:ara} is channel uniformity.  The second is that calibration pulse events resemble Askaryan pulses.  That is the purpose of the calibration pulser design \cite{allison2012design-cd3,allison2022lowthreshold-795}.  The pulser waveform is transmitted through an ARA dipole, propagates through ice to the A5 trigger string, and is converted to the observed traces.  This setup, however, does not alter $f_0$ or $\gamma$ in $r(t)$.  First, the distance from calibration pulser to trigger string is small compared to the RF attenuation length in ice, so the effect of frequency-dependent attenuation on the traces is negligible.  Second, the double convolution typically used to model transmission of an original calibration pulse through a transmitter to receiving antennas (see Eq. 9 of \cite{barwick2015timedomain-461}) does not alter $f_0$ because the calibration pulser antenna and the trigger antennas have the same dimensions, fixing $f_0$.  Finally, the envelope of the RF channel response $r(t)$ is modeled with an exponential, and the double convolution of two exponentials yields a function proportional to time multiplied by an exponential with the same $\gamma$ value.

Given these assumptions, we measured $f_0$ and $\gamma$ for the A5 trigger string using a data set of events triggered by the A5 calibration pulser.  Given that the lower edge of the RF channels is of order $\approx 0.1$ GHz, and that calibration pulser events are of order $\lesssim 100$ ns wide, we hypothesized that $f_0 \approx 0.1$ GHz and $\gamma \approx 0.01$ GHz.  Given these initial guesses, we calculated the CSW for each event, and maximized the correlation coefficient for each CSW, $\rho$, by tuning $\sigma_t$.  We scanned the $[f_0,\gamma]$ parameter space until the global average $\rho_{\rm ave}$ was maximized.  We found that $\rho_{\rm ave} = 0.79\pm 0.02$ was maximized when $f_0 = 0.19\pm 0.01$ GHz, and $\gamma = 0.02\pm 0.01$ GHz.  Though $\sigma_t$ was allowed to vary for each CSW, the maximum $\rho$ was reached when $\sigma_t = 0.6\pm 0.1$ for each CSW individually.  This is encouraging, for the calibration pulser properties should remain constant.  These results are summarized in Tab. \ref{tab:cal}.

\begin{table}
\centering
\begin{tabular}{| c | c | c |}
\hline
\textbf{ID} & \textbf{$\sigma_t$ (ns)} & $\rho$ \\ 
1995-0039 & $0.6\pm 0.1$ & 0.795 \\
2250-1064 & $0.6\pm 0.1$ & 0.802 \\
2250-1072 & $0.6\pm 0.1$ & 0.767 \\
2250-1081 & $0.6\pm 0.1$ & 0.776 \\
2250-1088 & $0.6\pm 0.1$ & 0.810 \\
2250-1107 & $0.6\pm 0.1$ & 0.816 \\
2250-1119 & $0.6\pm 0.1$ & 0.770 \\ \hline
& \textbf{Average:} & $0.79\pm 0.02$ \\ \hline \hline
\textbf{Best fit $f_0$, $\gamma$} [GHz] & $0.19\pm 0.01$ & $0.02\pm 0.01$ \\
\hline
\end{tabular}
\caption{\label{tab:cal} Results for $f_0$ and $\gamma$ from calibration pulser data.}
\end{table}

Several points about the calibration pulser events are worth mentioning.  As mentioned previously, the received voltage traces are technically the derivative of the original pulse, convolved twice with the effective height of the RF dipole (Eq. 9, \cite{barwick2015timedomain-461}).  For identical RF dipole transmitter and receiver, however, the parameter $f_0$ remains constant from transmitter to receiver because $f_0$ is set by the antenna dimensions.  Thus, fitting our model of the Askaryan pulse convolved with a single copy of the RF dipole response function will still oscillate with the correct frequency, $f_0$.  As a check, we fit the UHECR data while scanning $\sigma_t$, $f_0$, and $\gamma$ as free-parameters.  The result for $f_0$ was $0.19\pm 0.01$ GHz, remarkably consistent with the result in Tab. \ref{tab:cal}.  The result for $\gamma$ was $0.015\pm 0.02$ GHz, also consistent with Tab. \ref{tab:cal}.  The larger fractional error in $\gamma$ occurs for two reasons.  First, $\sigma_t$ can fluctuate in UHECR events but not calibration pulser events.  Second, the UHECR events contain small reflected pulses that alter $\gamma$ if it is allowed to vary event to event, rather than correctly treating it as a constant.

After measuring $f_0$ and $\gamma$, we proceeded to analyze the UHECR candidate voltage traces.  We computed the CSW for each UHECR candidate event, and maximized $\rho$ by only tuning $\sigma_t$, given our results for $f_0$ and $\gamma$ in Tab. \ref{tab:cal}.  The CSWs were found in most cases to contain a small reflected pulse.  In this context, a reflected pulse is a delayed copy of the primary pulse with a smaller amplitude.  For each UHECR candidate CSW, we performed a fit assuming two copies of our model, with widths $\sigma_{t,1}$ and $\sigma_{t,2}$, and time-delay between pulses $\Delta t$.  Initially, we set the amplitude of the reflection to 10\% relative to the primary pulse.  After obtaining the best-fit results for $\sigma_{\rm t,1}$, $\sigma_{\rm t,2}$, and $\Delta t$, we then tuned the relative amplitude of the reflection to minimize the power difference $\Delta P$ between CSW and model.  We define the $\Delta P$ between waveform data samples $d_i$ and the model waveform samples $m_i$ like

\begin{equation}
\Delta P = \sum_i (d_i - m_i)^2
\end{equation}

\noindent
The definition is loosely based on the power of an AC voltage, like $P = V^2/R$.  In our best fits for CSWs, we quote the reflection amplitude that minimizes $\Delta P$.

\begin{table}
\centering
\begin{tabular}{| c | c | c | c | c | c |}
\hline
\textbf{ID} & \textbf{$\sigma_{t,1}$ (ns)} & \textbf{$\sigma_{t,2}$ (ns)} & \textbf{$\Delta t$ (ns)} & \textbf{Rel. amp.} & \textbf{$\rho$} \\
1915-26288 & $0.6\pm 0.1$ & $2.0\pm 0.1$ & $25.25\pm 0.25$ & $0.08\pm 0.01$ & 0.86 \\
1957-13330 & $0.5\pm 0.1$ & $1.8\pm 0.1$ & $11.5\pm 0.25$ & $0.09\pm 0.01$ & 0.74 \\
2171-31805 & $0.6\pm 0.1$ & $1.9\pm 0.1$ & $25.25\pm 0.25$ & $0.09\pm 0.01$ & 0.81 \\
2250-20189 & $0.7\pm 0.1$ & $2.2\pm 0.1$ & $24.75\pm 0.25$ & $0.09\pm 0.01$ & 0.81 \\
2352-85489 & $0.7\pm 0.1$ & $2.1\pm 0.1$ & $24.75\pm 0.25$ & $0.10\pm 0.01$ & 0.84 \\
2375-17342 & $0.6\pm 0.1$ & $1.9\pm 0.1$ & $19.0\pm 0.25$ & $0.09\pm 0.01$ & 0.73 \\
2529-09767 & $0.7\pm 0.1$ & $1.9\pm 0.1$ & $25.5\pm 0.25$ & $0.11\pm 0.01$ & 0.82 \\
2716-58611 & $0.8\pm 0.1$ & -- & -- & -- & 0.84 \\
2782-00106 & $0.7\pm 0.1$ & $2.1\pm 0.1$ & $24.5\pm 0.25$ & $0.11\pm 0.01$ & 0.84 \\
2955-47449 & $0.5\pm 0.1$ & $1.6\pm 0.1$ & $25.0\pm 0.25$ & $0.03\pm 0.01$ & 0.69 \\
2961-98361 & $0.6\pm 0.1$ & $1.9\pm 0.1$ & $24.25\pm 0.25$ & $0.05\pm 0.01$ & 0.80 \\
2978-29412 & $0.5\pm 0.1$ & $1.7\pm 0.1$ & $25.5\pm 0.25$ & $0.05\pm 0.01$ & 0.82 \\
3352-89556 & $0.5\pm 0.1$ & $1.8\pm 0.1$ & $25.5\pm 0.25$ & $0.05\pm 0.01$ & 0.81 \\
\hline
\end{tabular}
\caption{\label{tab:results} Results for the primary pulse width $\sigma_{\rm t,1}$, reflected pulse width $\sigma_{\rm t,2}$, delay between primary and reflected pulses $\Delta t$, reflected pulse amplitude (relative to primary), and correlation coefficient $\rho$.}
\end{table}

\begin{table}
\centering
\begin{tabular}{| c | c | c | c |}
\hline
\textbf{ID} & \textbf{$\sigma_{t,1}$ (ns)} & \textbf{$\rho$} & \textbf{$\Delta \rho$} \\
1915-26288 & $0.5\pm 0.1$ & 0.83 & -0.03 \\
1957-13330 & $0.5\pm 0.1$ & 0.66 & -0.08 \\
2171-31805 & $0.5\pm 0.1$ & 0.775 & -0.035 \\
2250-20189 & $0.6\pm 0.1$ & 0.80 & -0.01 \\
2352-85489 & $0.6\pm 0.1$ & 0.83 & -0.01 \\
2375-17342 & $0.5\pm 0.1$ & 0.69 & -0.04 \\
2529-09767 & $0.6\pm 0.1$ & 0.80 & -0.02 \\
2716-58611 & $0.8\pm 0.1$ & 0.84 & -- \\
2782-00106 & $0.6\pm 0.1$ & 0.815 & -0.025 \\
2955-47449 & $0.6\pm 0.1$ & 0.68 & -0.01 \\
2961-98361 & $0.6\pm 0.1$ & 0.79 & -0.01 \\
2978-29412 & $0.5\pm 0.1$ & 0.78 & -0.04 \\
3352-89556 & $0.5\pm 0.1$ & 0.78 & -0.03 \\
\hline
\end{tabular}
\caption{\label{tab:results2} Same as Tab. \ref{tab:results}, but with no reflected pulse in the model. The term $\Delta \rho$ refers to the decrease in correlation coefficient listed in Tab. \ref{tab:results}.}
\end{table}

\begin{figure*}
\centering
\includegraphics[width=0.49\textwidth]{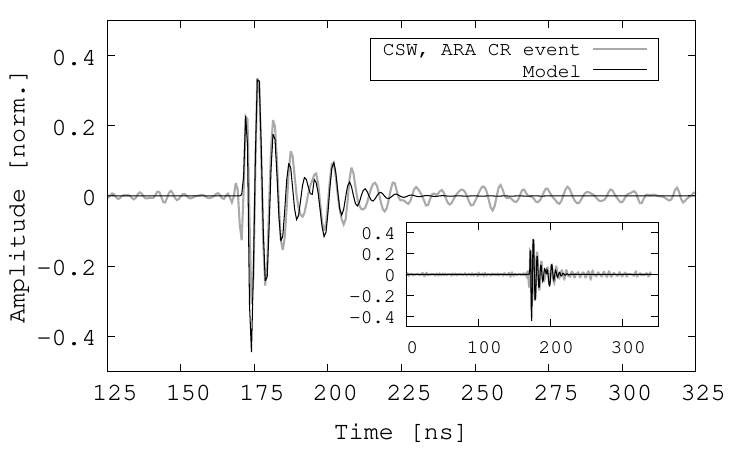}
\includegraphics[width=0.49\textwidth]{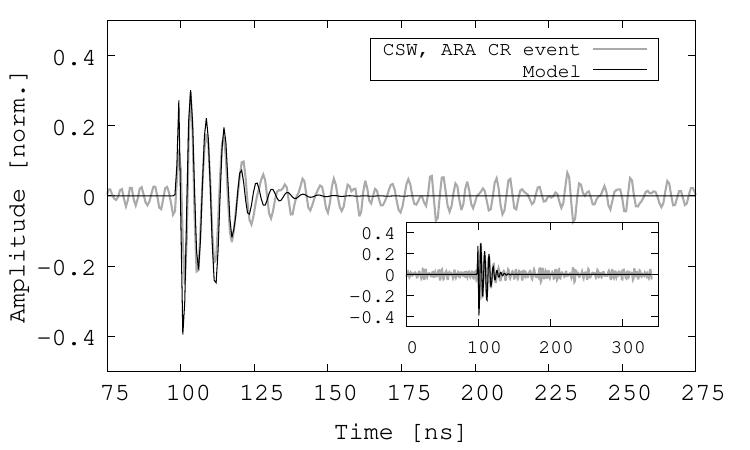}
\includegraphics[width=0.49\textwidth]{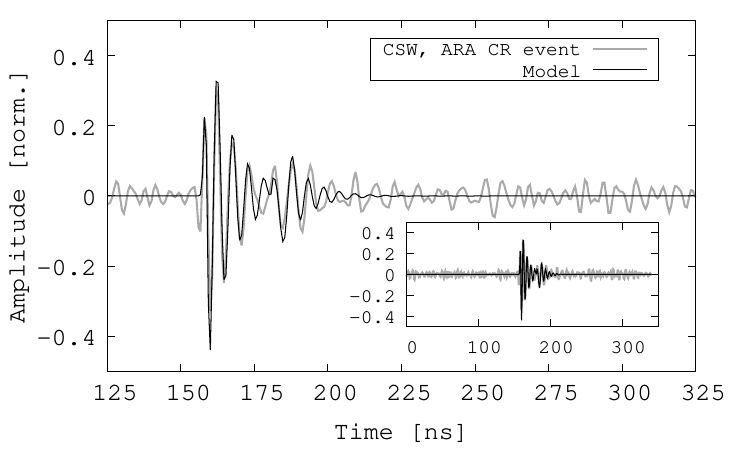}
\includegraphics[width=0.49\textwidth]{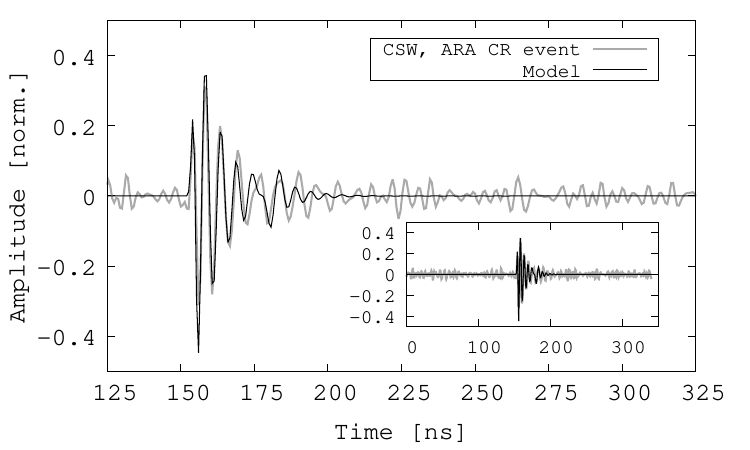}
\caption{\label{fig:1} UHECRs, by ID: (top left) 1915-26288, (top right) 1957-13330, (bottom left) 2171-31805, (bottom right) 2250-20189.  The insets of all graphs show the entire (normalized) raw voltage trace, for transparency.} 
\end{figure*}

\begin{figure*}
\centering
\includegraphics[width=0.49\textwidth]{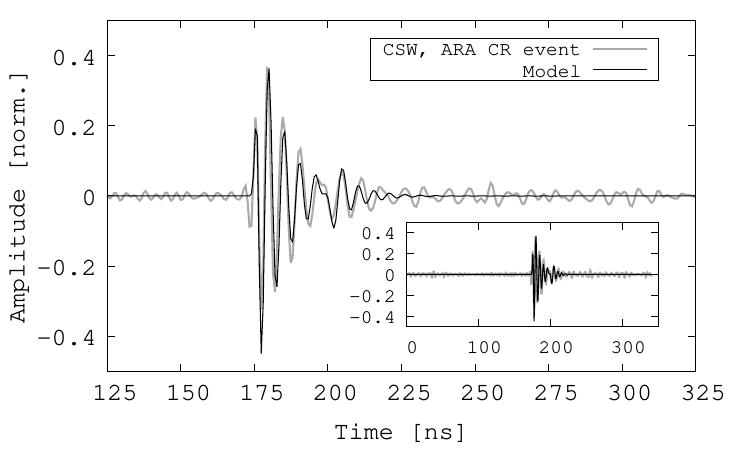}
\includegraphics[width=0.49\textwidth]{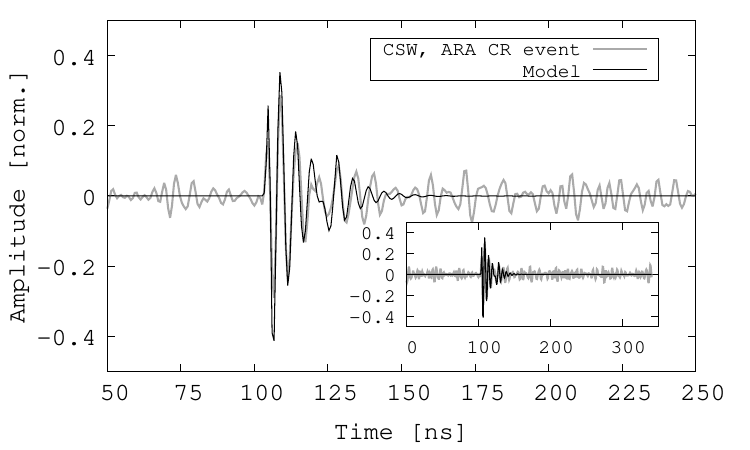}
\includegraphics[width=0.49\textwidth]{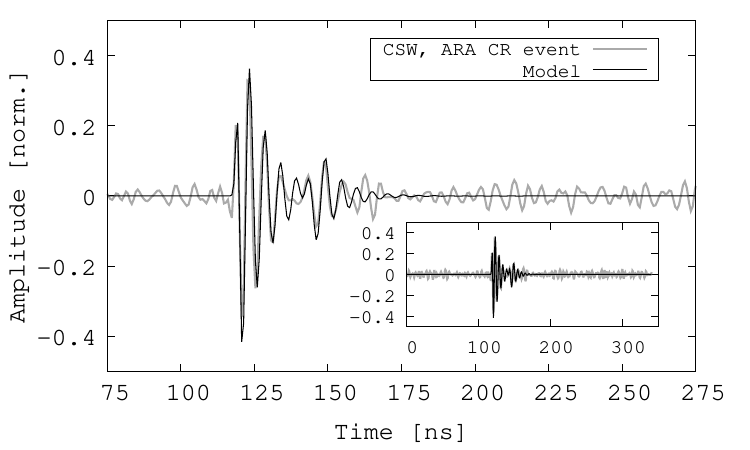}
\includegraphics[width=0.49\textwidth]{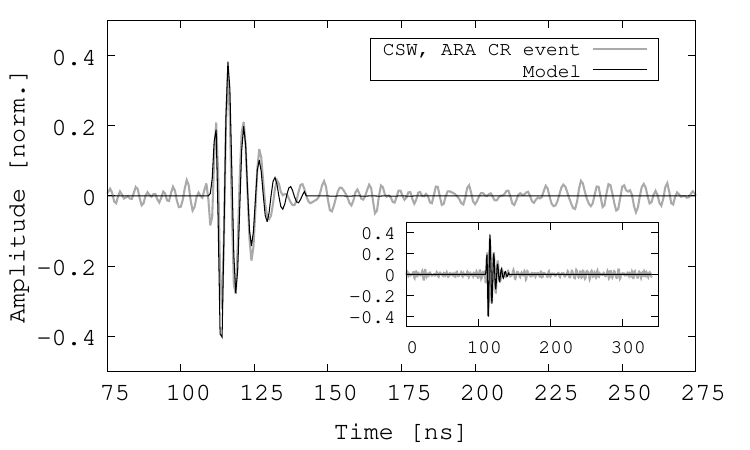}
\caption{\label{fig:2} UHECRs, by ID: (top left) 2352-85489, (top right) 2375-17342, (bottom left) 2529-09767, (bottom right) 2716-58611.  The insets of all graphs show the entire (normalized) raw voltage trace, for transparency.}
\end{figure*}

\begin{figure*}
\centering
\includegraphics[width=0.49\textwidth]{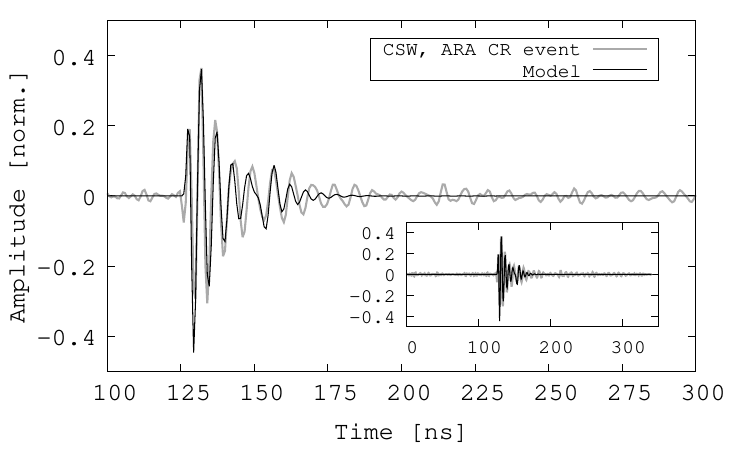}
\includegraphics[width=0.49\textwidth]{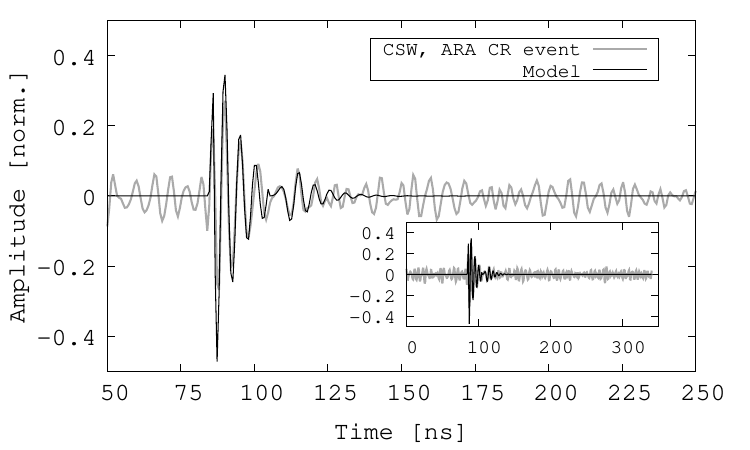}
\includegraphics[width=0.49\textwidth]{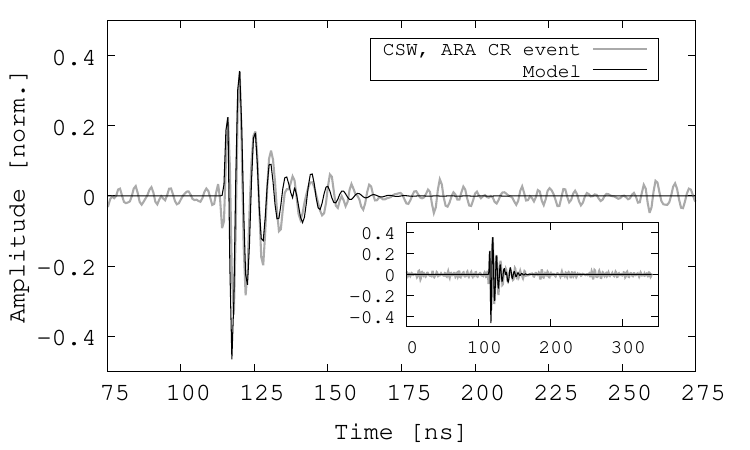}
\includegraphics[width=0.49\textwidth]{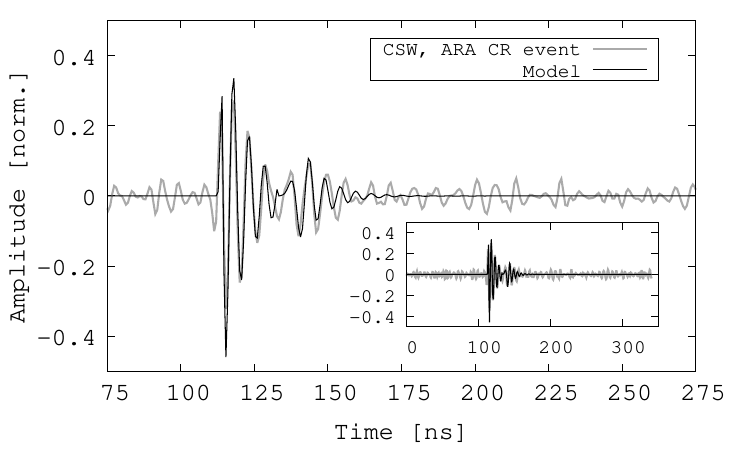}
\caption{\label{fig:3} UHECRs, by ID: (top left) 2782-00106, (top right) 2955-47449, (bottom left) 2961-98361, (bottom right) 2978-29412.  The insets of all graphs show the entire (normalized) raw voltage trace, for transparency.}
\end{figure*}

\begin{figure*}
\centering
\includegraphics[width=0.49\textwidth]{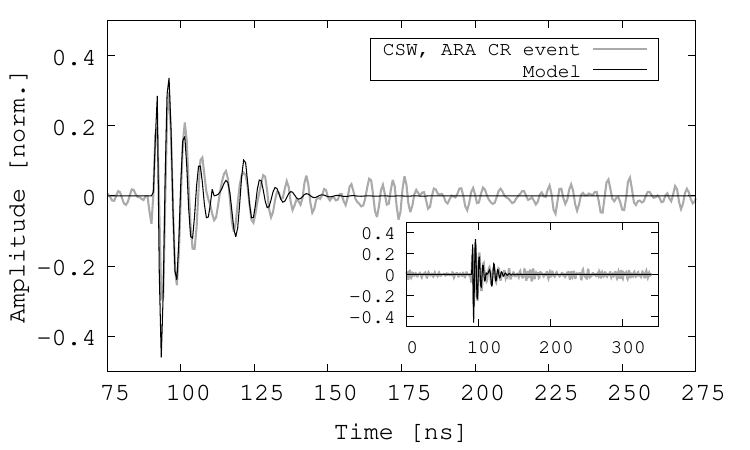}
\caption{\label{fig:4} UHECRs, by ID: 3352-89556.  The inset of the graph shows the entire (normalized) raw voltage trace, for transparency.}
\end{figure*}

The best-fit analytic traces and UHECR candidate CSWs are shown in Figs. \ref{fig:1}-\ref{fig:4}, and the numerical results for $\sigma_{\rm t,1}$, $\sigma_{\rm t,2}$, $\Delta t$, relative amplitude, and $\rho$ are shown in Tab. \ref{tab:results}.  For transparency, the inset graphs in all figures show the full CSW waveform from each event.  The same set of results, with relative amplitude set to zero, are shown in Tab. \ref{tab:results2}.  Reconstructed primary pulse widths, $\sigma_{\rm t,1}$, tend to be $<1$ ns, while reflected pulse widths, $\sigma_{\rm t,2}$, tend to be $>1$ ns.  The reflected pulses are delayed by $\approx 25$ ns, with two exceptions.  For event ID 1957-13330, the reflected pulse occurs within the envelope of the primary pulse, altering the total envelope (see Fig. \ref{fig:1} top right).  For event ID 2716-58611, there is no reflected pulse that can be fit by our algorithm.  The errors on $\sigma_t$, $\Delta t$, and relative amplitude are limited by our scan resolution.  Note that the sampling rate for the traces was 1.5 GHz.  Because $1/1.5$ GHz $\approx 0.67$ ns, we do not trust our timing precision far below 0.67 ns.

The results for $\sigma_{\rm t,1}$ in Tab. \ref{tab:results2} are similar to those in Tab. \ref{tab:results}.  For all cases the pulse width is within 0.1 ns of the result from Tab. \ref{tab:results} yielding correlation coefficients that are smaller by 1.25\% to 12\% relative to the results in Tab. \ref{tab:results}.  Thus, the reflected pulse in the model improves the fit, but not to an extent that changes the interpretation of the event as a UHECR.  For example, the authors of HH2026 \cite{hanson2026complex-a31} showed that UHE-$\nu$ events from an ARA-like detector would produce a distribution of correlation coefficients peaked at $\rho \approx 0.9$ but spread between 0.4 and 1.0.  If adding or removing the reflected pulse from the fit reduced $\rho$ below 0.4 this would have warranted a shift in interpretation. Because this is not the case, we chose not to interpret the origin of the reflections. Note that reflections are common when receiving antenna are located near other antennas, metal conductors, or RF cables.

A final point regarding the fits for $\sigma_t$ is worth mentioning.  In Eq. \ref{eq:sr2}, $\sigma_t$ is fit as a free parameter.  In our previous work, we derived the relationship between the instantaneous charge distribution (ICD) form factor of the particle cascade, the spectral cutoff frequency of the Askaryan radiation, and $\sigma_t$ \cite{10.1016/j.astropartphys.2017.03.008,PhysRevD.105.123019}.  The model was derived under the assumption that the density and index of refraction of the solid medium are constants.  Switching the medium from solid ice to snow changes the ICD, which in turn changes the cutoff frequency.  The new cutoff frequency, in turn, affects $\sigma_t$.  Our model still fits the data, however, indicating that the results for $\sigma_t$ correspond to UHECR cascades within snow.  Since the UHECR shower cascades in our analysis occur over $\mathcal{O}(10)$ meters in snow, we assume a constant density, albeit a different one from deep ice.  For examples of RF index of refraction data in polar ice shelves and ice sheets, see \cite{10.3189/2015jog14j214,Barwick:2018497}.

Future fits to UHE-$\nu$ candidate events from deep ice might differ from these results for two reasons.  First, the fits could improve because the entire UHE-$\nu$ cascade begins and ends in the ice, rather than an unknown portion of it in our current case.  Second, the fit could worsen because our model does not account for frequency-dependent RF attenuation in ice.  The latter should be a minor effect, however, given the analysis in Sec. 5 of \cite{10.1016/j.astropartphys.2014.09.002}, Figs. 12 and 13.

\section{Conclusion}
\label{sec:conc}

\begin{figure*}
\centering
\includegraphics[width=0.67\textwidth]{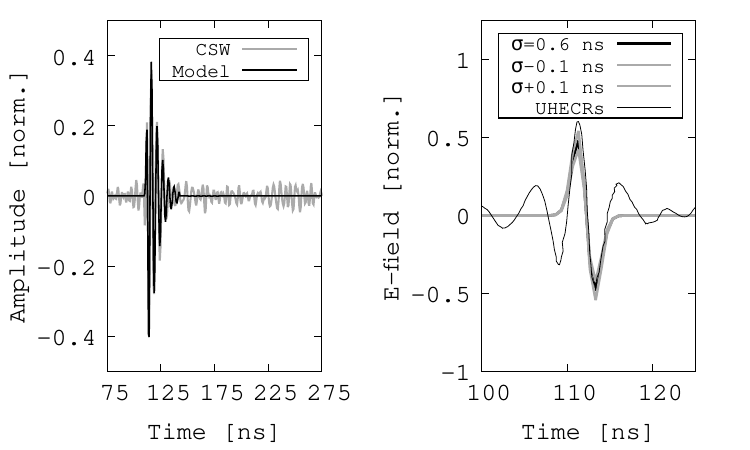}
\caption{\label{fig:ex} (Left) The results from Fig. \ref{fig:2} bottom right, event ID 2716-58611. (Right) Equation \ref{eq:s}, evaluated with the reconstructed $\sigma_t$ from Tabs. \ref{tab:results} and \ref{tab:results2}, and $E_0 = 1$.  Thick black line: $\sigma_t = 0.6$ ns.  Thick gray lines: $\sigma_t = 0.7$ ns and $\sigma_t = 0.5$ ns.  Thin black line: average data from UHECR $\vec{E}$-fields, adapted from Fig. 2 of \cite{xwqy-yzrk}.}
\end{figure*}

An example of what the analysis reveals about the UHECR candidate data is shown in Fig. \ref{fig:ex}.  In Fig. \ref{fig:ex} (left) the CSW fit from Fig. \ref{fig:2} bottom right, event ID 2716-58611, is shown.  This CSW does not contain a reflected pulse, and the correlation between model and data was 0.84.  The reconstructed $\sigma_t$ was $0.6 \pm 0.1$ ns.  Knowing $\sigma_t$ allows us to graph the reconstructed $\vec{E}$-field, normalized with $E_0=1$ in Eq. \ref{eq:s}.  The field is graphed in Fig. \ref{fig:ex} (right), for $\sigma_t = 0.6$ ns, accounting for the uncertainty in $\sigma_t$.  The reconstructed $\vec{E}$-field from our analysis matches the central portion of the blue data shown in Fig. 2 of \cite{xwqy-yzrk}.  The data was taken from Fig. 2 of \cite{xwqy-yzrk}, smoothed with a 10-sample running average filter, and inverted.  Our analytic reconstruction technique (fitting Eq. \ref{eq:sr2} to CSWs from raw, observed voltage traces) produces an $\vec{E}$-field field that matches the deconvolution used to produce the reconstructed $\vec{E}$-field in Fig. 2 of \cite{xwqy-yzrk}.  Note that we are not \textit{fitting} our model to the data from Fig. 2 of \cite{xwqy-yzrk}, but merely plotting it on the same axes.

Deconvolution algorithms used to produce the $\vec{E}$-field in Fig. 2 of \cite{xwqy-yzrk} usually involve three stages.  The first step is to measure the impulse response $r(t)$ of the entire RF detection chain in an anechoic chamber or RF bench testing environment.  Once the data from the deployed detector is collected, the second stage involves filtering it for noise.  Thermal noise has a detrimental effect on de-convolution algorithms \cite{jansson2012deconvolution}.  Other forms of noise require techniques like notch filtering, which can cause ringing oscillations like those observed in Fig. 2 of \cite{xwqy-yzrk}.  The final step is to use techniques like forward-folding or the convolution theorem to calculate the $s(t)$ that causes $s(t) * r(t)$ to match the data.  The three-stage process  contains opportunities for error propagation, and the match in Fig. \ref{fig:ex} (right) demonstrates the deconvolution algorithms used to produce the $\vec{E}$-field from Fig. 2 of \cite{xwqy-yzrk} agree with our analytic model.

Further, measurements of $\sigma_t$ from UHECR candidates can be used to calculate parameters of UHECR cascade observation.  Consider Eq. 5 of \cite{hanson2026complex-a31}:

\begin{equation}
a\Delta\theta = \frac{c \sigma_t}{\sin\theta_{\rm C}} \label{eq:uncert}
\end{equation}

\noindent
In Eq. \ref{eq:uncert}, $a$ is a measure of the longitudinal length of the in-ice cascade in meters, $\Delta\theta$ is $\theta - \theta_{\rm C}$, where $\theta$ is the viewing angle with respect to the cascade axis, and $\theta_{\rm C}$ is the Cherenkov angle, $c$ is the speed of light in ice, and $\sigma_t$ is the pulse width.  The authors of \cite{PhysRevD.105.123019} calculated for 10-100 PeV cascades, $a \approx 5$ m.  Using $\sigma_t = 0.6$ ns, and $c = 0.3/1.4$ m ns$^{-1}$ and $\theta_{\rm C} = 0.775$ radians for RF waves in snow, we find $\Delta\theta \approx 2.1$ degrees.  This result aligns with observed UHECR candidate power spectra shown in Fig. 2 of \cite{xwqy-yzrk}.  The red band in Fig. 2 of \cite{xwqy-yzrk} is calculated assuming $\Delta\theta \leq 3$ degrees.  Assuming that the UHECR cascade energy was between 10 and 100 PeV gives the correct range for $\Delta\theta$.

There are some logical steps for this work in the future. First, a separate fit for $\Delta\theta$ would allow the measurement of the $a$ parameter, which is tied to the UHECR cascade energy. This parameter $\Delta\theta$ may be constrained by using the spectral cutoff of the voltage traces, which are inversely correlated to $\Delta\theta$ \cite{10.1016/j.astropartphys.2017.03.008}.  Second, it is interesting to note that our model for $s(t)$ in Eq. \ref{eq:s} corresponds to the \textit{off-cone} ($\theta\neq \theta_{\rm C}$) result from HH2022 \cite{PhysRevD.105.123019}.  The authors of \cite{PhysRevD.105.123019} also include an \textit{on-cone} version of $s(t)$ that is used when $\Delta\theta = 0$, and can be used as $\Delta\theta\to 0$.  The on-cone model could provide a better fit for events with $\Delta\theta \lesssim 1$ degree \cite{PhysRevD.105.123019}.  Finally, it is our hope that one day our model will be used to fit UHE-$\nu$ candidates, in addition to UHECR candidates.

\section{Acknowledgements}

We would like to thank the ARA collaboration for sharing the UHECR candidate data, and the corresponding calibration pulser data.  We note that the data used in this analysis have been publicly released on the Zenodo platform (\url{https://zenodo.org/records/19578836}).  Finally, we would like to thank ARA collaboration members for many helpful comments and suggestions.

\bibliography{apssamp}

\end{document}